# Bias Dependence and Electrical Breakdown of Small Diameter Single-Walled Carbon Nanotubes

R. V. Seidel*, A. P. Graham, B. Rajasekharan, E. Unger, M. Liebau, G. S. Duesberg, F. Kreupl, W. Hoenlein

Infineon Technologies AG, Corporate Research, 81730 Munich, Germany

*Corresponding author. Phone: (+49) 89 234-52755, E-mail: robert.seidel@infineon.com

**ABSTRACT**

The electronic breakdown and the bias dependence of the conductance have been investigated for a large number of catalytic chemical vapor deposition (CCVD) grown single-walled carbon nanotubes (SWCNTs) with very small diameters. The convenient fabrication of thousands of properly contacted SWCNTs was possible by growth on electrode structures and subsequent electroless palladium deposition. Almost all of the measured SWCNTs showed at least weak gate dependence at room temperature. Large differences in the conductance and breakdown behavior have been found for "normal" semiconducting SWCNTs and small band-gap semiconducting (SGS) SWCNTs.

# I. INTRODUCTION

The exciting electronic properties of single-walled carbon nanotubes have been the focus of research for some time.[1] In order to assess their potential for nanoelectronic applications, such as transistors or interconnects, it is necessary to obtain a more detailed understanding of their behavior at different biases and their electronic breakdown. The simultaneous synthesis of SWCNTs with different electronic properties is still the largest roadblock for the direct integration of SWCNTs into electronic devices. Semiconducting SWCNTs with a high on/off current ratio are necessary for transistors, whereas metallic nanotubes are required for interconnects. Further, when considering the behavior of metallic nanotubes it is important to distinguish between armchair SWCNTs, with truly metallic character and "metallic" SWCNTs with a small curvature-induced band-gap. Although some methods to separate metallic and semiconducting nanotubes have been suggested, they lack precision, have poor yield, and in some cases still require proof by electronic characterization of the separated material.[2-5]

Preferential breakdown is still the only straight-forward method to eliminate nanotubes that cannot be completely turned off at room temperature. The electrical breakdown of bundles of SWCNTs has, to some extent, been investigated previously.[6,7] Preferential breakdown can be achieved by depleting the semiconducting SWCNTs with an appropriate gate voltage before increasing the bias along the tubes until they burn through. Since as-grown semiconducting SWCNTs are usually p-type, a positive gate voltage ($V_g$) has to be applied. The current carrying capability of individual metallic SWCNTs has been found to be as high as 25 µA and the breakdown has been related to self-heating by electron-phonon scattering processes and thermal oxidation. Further, electrical breakdown of individual SWCNTs has been reported and employed to determine the proportion of semiconducting SWCNTs.[8]

The performance of high-current nanotubes transistors, recently presented by our group, still depends largely on the preferential electrical breakdown of the nanotubes that contribute to the metallic conduction.[9] Those transistors are built from the parallel arrangement of a large number of randomly grown SWCNTs. To optimize the properties of these transistors the electronic breakdown of SWCNTs with various electronic properties has to be understood. In this work a study of the room temperature transport characteristics of SWCNTs with small diameters will be presented.

## II. EXPERIMENTAL

Carbon nanotubes were grown by catalytic chemical vapor deposition at 700 °C between electrode structures consisting of a Ta/Co/Al/Ni multilayer defined with optical lithography (350 nm). P-type silicon with 200 nm thermally grown oxide was used as the substrate. A thin (< 0.3 nm) layer of Ni serves as the catalyst for SWCNT growth and methane was used as the carbon feedstock. The growth of SWCNTs at moderate temperature and on electrode structures has been discussed in detail elsewhere.[10,11] The density of nanotubes was adjusted so that on average only one SWCNT bridges the electrodes. In this way bundle formation can typically be excluded. The contacts to the nanotubes were improved by electroless Pd deposition at 50 °C and annealing at 400 °C in $N_2$ after the growth. For the electroless Pd deposition an alkaline tetraamminepalladium (II) chloride solution together with a reducing agent was used. This process is similar to the electroless nickel deposition first applied for CNTs by our group.[10,12] The electroless Pd deposition yields good electrical contacts to the nanotubes similar to those obtained with electroless Ni with the advantage of improved control over the deposition and higher corrosion resistance. Recently, it has been shown that Pd yields good contacts to SWCNTs because of its high work-function and good wetting properties.[13,14] For comparison, SWCNTs were also contacted by e-beam lithography and evaporation of a 30 nm Pd film in order to rule out that the electroless deposition process alters the properties of the nanotubes. No obvious difference was found in the performance of SWCNTs with evaporated or electroless Pd contacts.

The resulting field-effect-transistors were controlled using the Si substrate as a back-gate. The electronic measurements were performed using a Keithley 4200 semiconductor characterization system in an air environment and at room temperature. The measurement accuracy for small currents was limited by the measurement unit to around 10 pA. Scanning electron microscopy (SEM) and atomic force microscopy (AFM) studies were carried out using a LEO 1560 and a DI Dimension 5000, respectively.

## III. RESULTS AND DISCUSSION

It has been frequently observed that the growth temperature affects the diameter range of the SWCNTs.[15] Decreasing the growth temperature usually causes a reduction of the average diameters. We observed rather small diameters for the SWCNT grown at 700°C using the process described above. Raman spectroscopy revealed that the diameters of nanotubes grown on quartz substrates from thin Ni catalyst layers on Al layers under identical conditions are between 0.7 and 1.6 nm. AFM measurements on the measured transistors also confirmed this diameter range, but also showed that all of the longer nanotubes have small diameters (< 1.1nm). Only a few SWCNTs with diameters between 1.1 nm and 1.6 nm were found by AFM, remarkably all of them were too short to bridge electrodes with a larger separation ≥ 1 µm. Therefore, large diameter SWCNTs can be excluded from the discussion.

The structured growth of nanotubes coupled with electroless Pd deposition enabled the generation of a very large number of well-contacted SWCNTs without laborious e-beam lithography. Several hundred nanotube devices were measured in order to determine their properties and extract the general trends. The majority of the measured SWCNTs showed at least small gate dependence at room temperature, i.e. a current modulation of at least a factor of 3 while sweeping the gate voltage from –20 V to +20 V. In detail the nanotubes can be divided into three groups depending on the gate voltage dependence; 75 % to 80 % of the nanotubes showed pronounced semiconducting behavior with a current modulation of more than 5 orders of magnitude, whereas between 20 % and 25 % showed a modulation by a factor of 3 to 100, and less than about 2 % of the measured nanotubes showed considerably smaller gate dependencies of less than 30 %. These nanotubes are probably truly metallic SWCNTs with an armchair structure. Examples of the room temperature characteristics of the investigated small diameter SWCNTs will be discussed according to those three categories.

Starting with the metallic behavior, Fig. 1a shows the gate voltage ($V_g$) dependence of the drain current ($I_{ds}$) for two metallic SWCNTs grown between electrodes with 400 nm separation. Figure 1b shows the corresponding breakdown curves measured during the application of a positive gate voltage of +30 V. It is important to note that such metallic nanotubes were only found between electrodes with a rather small separation (400 – 500 nm) and were never observed between electrodes with larger separation ($\geq$ 1 µm). Therefore, we propose that the chirality of a specific nanotube influences the length via the growth rate. If there is such a dependence of the growth rate on the chirality, then this effect will become more pronounced at lower temperatures and might also explain why about 90 % of the SWCNTs grown by remote plasma enhanced CCVD at 600 °C are semiconducting.[8] In addition, there might also be a dependence of the growth speed on the diameter. This would explain why AFM measurements indicate that all of the longer nanotubes have very small diameters.

Figure 2 shows an example of the gate voltage dependence of the drain-source current ($I_{ds}$) of the SWCNTs belonging to the group made up of the 20 - 25 % of the nanotubes that show a much larger modulation than the metallic SWCNTs but cannot be fully turned-off at room temperature like the semiconducting SWCNTs. Their modulation ranges between 3 and about 100. These 20 – 25 % are most probably small band gap semiconducting (SGS) SWCNTS, with a curvature induced band gap,[16-19] consistent with theoretical and experimental expectations.[19] Figure 3a shows that the gate voltage dependence of the drain-source bias of these nanotubes has a pronounced dip around $V_g = 0$ V for low biases, indicative of a small band gap. Further, the transition from predominantly p-type in air to n-type in vacuum shown in Fig. 3b indicates that the transport characteristics depend on doping by adsorbed species that evaporate in vacuum. This is typical of SWCNTs with some semiconducting properties since the conduction in purely metallic tubes should not be affected by adsorbed species except, perhaps, to reduce the scattering length.

Figure 4a shows the logarithmic $I_{ds}$ vs. $V_g$ plots of the same device shown in Fig. 2 at different drain-source biases ($V_{ds}$). It can be seen that the modulation increases slightly at smaller biases. Figure 4b shows the dependence of $I_{ds}$ on the bias for the application of a constant negative gate voltage of -20 V. Breakdown occurs after a long saturation phase. Figure 4c shows the conductance derived from the data of Figure 4b. The conductance starts to decrease from the initial value of 45 µS with increasing bias until the nanotube fails by electrical breakdown at 3-4 µS. The high value of the conductivity compared to the semiconducting tubes shown later is probably a result of smaller contact resistances to the SGS nanotubes due to smaller Schottky barriers. The decrease of the conductance with increasing bias voltage is related to an increase of electron-phonon scattering and the corresponding self-heating. Figure 4d shows an SEM image of the same SWCNT after breakdown. The location of the break can be easily identified as the point at which differences in the charging characteristics lead to a contrast variation. For most of the nanotubes the breakdown occurred approximately half way between the contacts indicating that the breakdown is the result of bulk heating of the nanotube; the contacts act as heat sinks reducing the probability of breakdown in their vicinity.

The remaining 75 – 80 % of SWCNTs showed a much more pronounced semiconducting behavior than the SGS SWCNTs with an on/off current ratio of typically more than 5 orders of magnitude. Figure 5a shows the gate voltage dependence of the semiconducting SWCNT shown in Fig. 5c together with the bias voltage dependence for a gate voltage $V_g$ of –20 V in Fig. 5b. This SWCNT shows a maximum conductance in the on-state of only 2 µS at low bias, although values of up to 6 µS have also been observed for other semiconducting SWCNTs. This is considerably lower than the maximum conductance of the metallic SWCNTs and the SGS SWCNTs with around 10 – 50 µS and the semiconducting SWCNTs with larger diameters as shown by Javey et al.[13] This substantial difference can be attributed to large Schottky barriers, which result from the large band gap of small diameter (< 1.1 nm) nanotubes and to a considerable degree from back scattering of the carriers.[20,21] Higher saturation currents can only be achieved for semiconducting SWCNTs with lengths below the estimated mean free path of around 100 - 300 nm or with larger tube diameters.[13,22]

The conductance of the SWCNT shown in Fig. 5c was measured as a function of the drain-source bias voltage in the on-state ($V_g = -20$ V). The conductance decreases monotonically until breakdown at $V_{ds} = 21$ V (Fig. 5b). Figure 5d presents a magnified SEM image of the location of the breakdown, which is surrounded by a round halo. The gate voltage dependence curves for a different semiconducting SWCNT are shown in Fig. 6a together with the dependence of the conductance on the bias in Figure 6b. A significant increase of the conductance at small biases can be observed. For this nanotube the initial conductivity increase is much more pronounced than for the other semiconducting SWCNT shown in Fig. 5b. However, this initial increase of the conductance is characteristic for all of the measured semiconducting SWCNTs, in contrast to the SGS SWCNTs where the maximum conductance is always observed at minimum drain-source bias. This increase of the conductance is related to tunneling through the Schottky barriers, which depends exponentially on the bias. The maximum of the conductance appears usually at a bias of 0.5 to 1 V for nanotubes with lengths between 0.4 and 2 µm. The exact length dependence could not be determined since it is strongly affected by the precise electronic structure (e.g. band gap and defects) of the individual nanotubes.

Breakdown of the semiconducting SWCNTs occurs at drain-source biases between 5 and 25 V depending on the contact resistances, nanotube length, applied gate voltage, heat transfer to the contacts and its structural perfection. Since electrical transport in longer semiconducting SWCNTs is diffusive their breakdown will primarily depend on the electric field along the tube. The breakdown is connected with large heat dissipation, which is already a crucial issue in current ICs with a high integration density. Therefore, nanotube transistors should be operated in a range where the conductance is maximum and the bulk heating small. That means they should be operated in the bias range before current saturation or they have to be sufficiently short to allow ballistic transport.

The semiconducting SWCNTs shown in Figure 5 and 6 have been burned-through in the on-state ($V_g = -20$V). In contrast, Figure 7 shows the breakdown of a 4 µm long and approximately 0.7 nm diameter semiconducting SWCNT burned-through in the off-state at a gate voltage $V_g = +20$V. As the bias increases the influence of the gate is diminished and the conduction increases. Breakdown occurs at about 25 V and a maximum current of 5 µA, i.e. 200 nS similar to the nanotubes shown in Figs. 5 and 6.

The previously discussed breakdown of SWCNTs has been recently used to estimate the proportion of metallic and semiconducting SWCNTs.[8] At first glance this method would seem to be a reliable way to count the number of metallic and SGS SWCNTs since they usually show smooth $I_{ds}$ vs. $V_{ds}$ curves with well defined steps during breakdown. In contrast the semiconducting SWCNTs often exhibit strong variations of the current in the saturation regime independent of the applied gate voltage, as illustrated in Fig. 8. The resulting multiple peaks might be misleading if their number is used to estimate the number of SWCNTs. The current oscillations are also detrimental to the smooth operation of SWCNT transistors at current saturation and high drain-source biases.

The presence of SGS SWCNTs also has strong implications for the operation of high-current transistors constructed from many nanotubes connected in parallel, which was recently presented by our group.[9] Initially, it was assumed that the off-current is predominantly due to nanotubes with almost no gate dependence whereas the present evidence indicates that this is not correct. Further, the on-current prior to burn-through of the nominally metallic tubes is the sum of the on-currents of the semiconducting SWCNTs, the SGS SWCNTs and the very few genuinely metallic SWCNTs. Thus, the absolute difference between the on and off currents cannot be maintained after the SGS SWCNTS have been destroyed by electrical breakdown since the on-currents of individual SGS SWCNTs are usually much larger than those of semiconducting SWCNTs, as discussed above. This is due to the lower Schottky barrier heights resulting from the smaller energy gaps of the SGS nanotubes.

An example of the effect of the SGS nanotubes on the transistor characteristics is illustrated in Figure 9. In this case at least three SWCNTs were grown between the electrodes, a number that allows the individual contributions of each tube to be evaluated by breakdown at high bias. The top curve (sgs1+sgs2+sc) in Fig. 9a shows the gate voltage dependence prior to breakdown and the lower curves (sgs2+sc and sc) following the elimination of two of the SWCNTs at a gate voltage of $V_g = +20$ V. The contribution of each tube to the conduction characteristics can, therefore, be obtained by evaluating the difference between the successive curves, as shown in Fig. 9b. This shows that the two tubes that have been eliminated by electrical breakdown are small band gap SWCNTs (sgs1 and sgs2 in Fig. 9b) by comparison with the curves shown in Figs. 2, 3 and 4a. The remaining nanotube is semiconducting (sc) by comparison with Figs. 5a, 6a and 7a.

The initial on and off-currents in Fig. 9a are 30 µA and 7 µA, respectively, corresponding to current difference of 23 µA. Hence, if pure metallic SWCNTs were responsible for the off-current of 7 µA, then the on-current after elimination of those tubes would be 23 µA. However, since the nanotubes that are eliminated are small band-gap SWCNTs, and not metallic tubes, the on-current of the remaining semiconducting nanotube is 7 µA, only about one quarter of the initial value.

## IV. CONCLUSIONS

In summary, the bias dependence and electrical breakdown of CCVD grown SWCNTs have been investigated. The diameters of the characterized nanotubes were in the range of 0.7 to 1.1 nm due to the rather low growth temperature and small catalyst layer thickness. About 98 % of the measured SWCNTs show a gate dependent current modulation of at least a factor of 3. Only 2 % of the nanotubes displayed pure metallic behavior and they were only found between contacts with a narrow separation. This suggests that the growth speed depends on the chirality. Small band gap semiconducting (SGS) SWCNTs were found to account for about 20 – 25 % of the total number of nanotubes. The small band gap of these tubes is predicted to derive from a curvature induced band bending, which results from the narrow diameter. The bias dependent conductance behavior of SGS SWCNTs and normal semiconducting SWCNTs are quite different. The semiconducting SWCNTs were found to have a conductance maximum at a bias of around 0.5 to 1 V, whereas an immediate decrease in conductance is observed for SGS SWCNTs. SGS nanotubes breakdown at currents of around 20 µA, on the other hand semiconducting SWCNTs with diameters < 1.1 nm and lengths longer than about 300 nm can usually withstand only 3 - 6 µA. These results are helpful in assessing SWCNTs with small diameters for future applications. The almost complete absence of metallic SWCNTs grown under the current CCVD conditions could be advantageous for transistor applications but detrimental for interconnects built from bundles of SWCNTs. Control of the small band gap semiconducting SWCNTs still poses a huge challenge, which might be solved by band gap opening as a result of chemical or mechanical modifications or sophisticated gate design.

**ACKNOWLEDGMENTS**

This work has been supported by the German Ministry of Science and Technology (BMBF) under Contract No. 13N8402.

**Figure 1.** $I_{ds}$ vs. $V_g$ plots (a) of two different metallic SWCNTs and the corresponding breakdown curves (b).

**Figure 2.** $I_{ds}$ vs. $V_g$ plot of a small band gap semiconducting SWCNT.

**Figure 3.** $I_{ds}$ vs. $V_g$ plots (a) of a small band gap semiconducting SWCNTs at different biases providing some evidence of the small band gap (dip). The performance of SGS SWCNTs is changed in vacuum from more p-type to more n-type (b). This is typical of SWCNTs with a semiconducting contribution since the conduction in purely metallic tubes should not be affected by adsorbed species.

**Figure 4.** $I_{ds}$ vs. $V_g$ plots (a) of a small band gap semiconducting SWCNT, the corresponding breakdown curve (b), and derived conductance plot (c). The SEM image in (d) shows the nanotube and the location of the breakdown.

**Figure 5.** $I_{ds}$ vs. $V_g$ curve ($V_{ds}$ = 1V) of a 1.5 μm long and approximately 0.8 nm diameter p-type SWCNT (a), conductance vs. $V_{ds}$ curve of the nanotube in the on-state ($V_g$ = -20 V) (b), and corresponding SEM images (c) and (d). The SEM image in (d) shows the location of the breakdown.

**Figure 6.** $I_{ds}$ vs. $V_g$ curve ($V_{ds}$ = 1V) of a 1 µm long p-type semiconducting SWCNT (a) and conductance vs. bias curve of the nanotube in the on-state ($V_g$ = -20 V). The strong shift towards negative $V_g$ is related to hysteresis. (b). An initial increase of the conductance can be observed at small biases. At larger biases the current saturates and the conductance decreases again until electrical breakdown occurs.

**Figure 7.** $I_{ds}$ vs. $V_g$ (a) and $I_{ds}$ vs. $V_{ds}$ plots (b) showing the semiconducting behavior and the electrical breakdown of a SWCNT (length about 4 µm; diameter approx. 0.7 nm). The $I_{ds}$ vs. $V_{ds}$ dependence was measured with an applied gate voltage of +20 V so that the tube is turned off at small biases. At high bias fields (> 15 V) the gate effect is suppressed and the tube is turned on.

**Figure 8.** $I_{ds}$ vs. $V_{ds}$ plot (a) showing the electrical breakdown of a single semiconducting SWCNT with $V_g$ = -20V. Strong oscillations of the current occur at saturation.

**Figure 9.** $I_{ds}$ vs. $V_g$ plots (a) showing the characteristics of a parallel arrangement of two SGS SWCNTs (sgs1 and sgs2) and at least one semiconducting SWCNT (sc). The curves in (a) show the total currents of the nanotubes in parallel in the initial state and after breakdown of each of the two SGS SWCNTs, whereas (b) shows the $I_{ds}$ vs. $V_g$ of the individual contribution of the nanotubes obtained by subtracting the data after each breakdown. The inset in (b) shows the electrical breakdown of the two SGS SWCNTs.

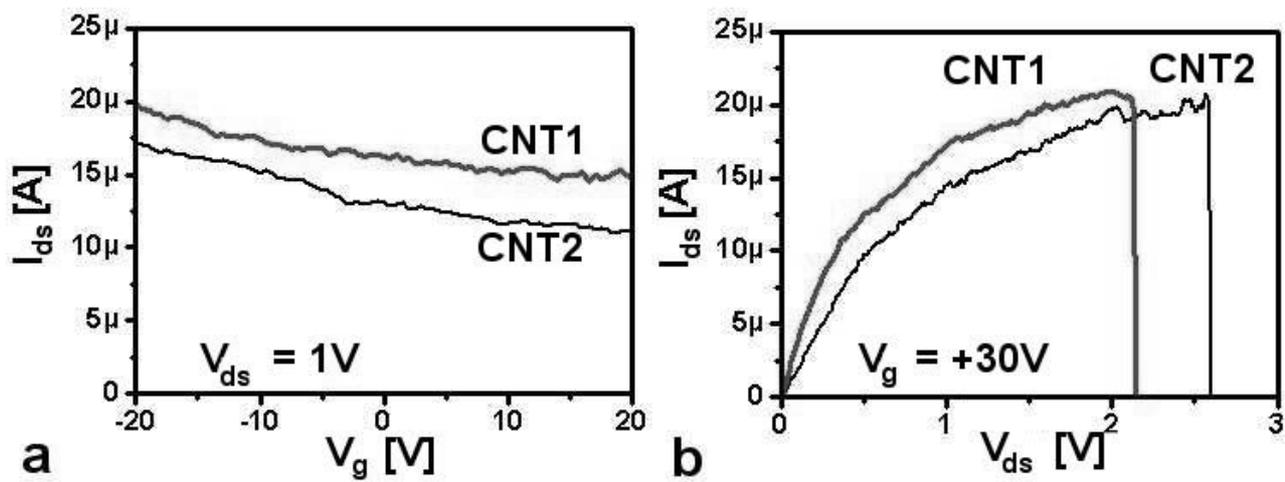

Figure 1

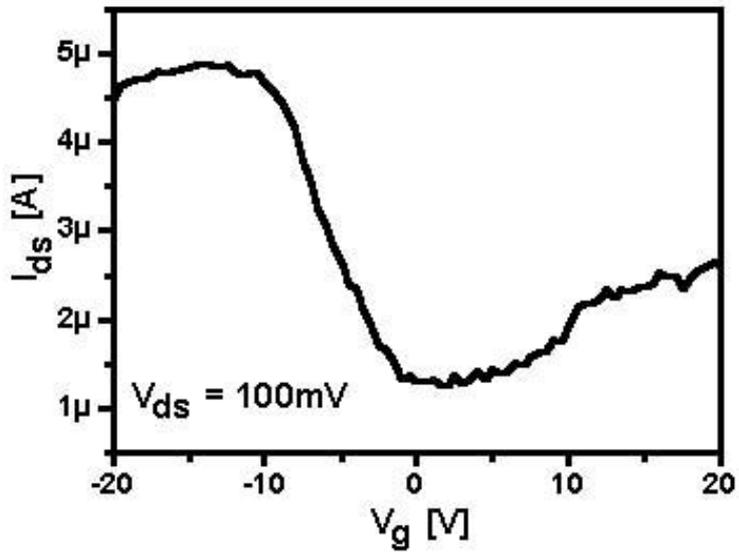

Figure 2

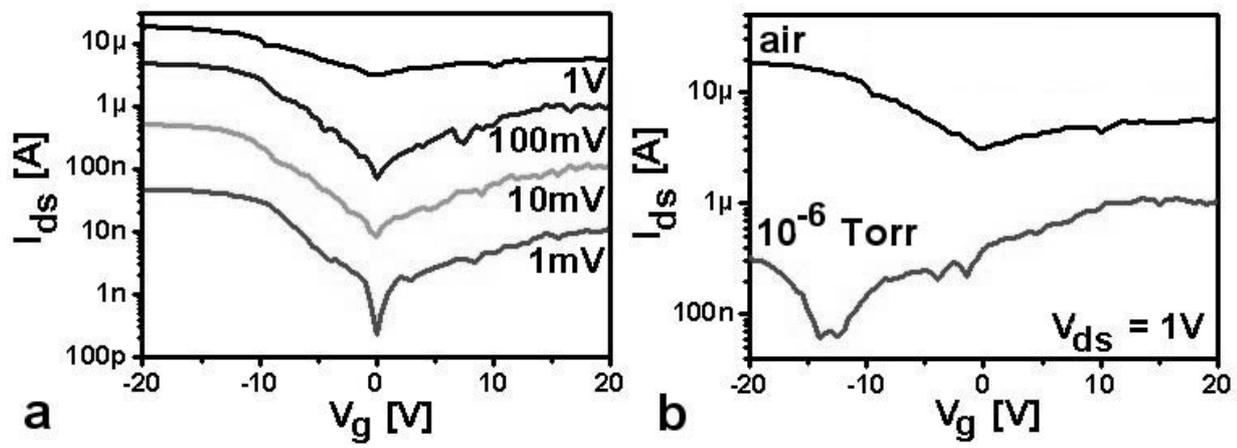

Figure 3

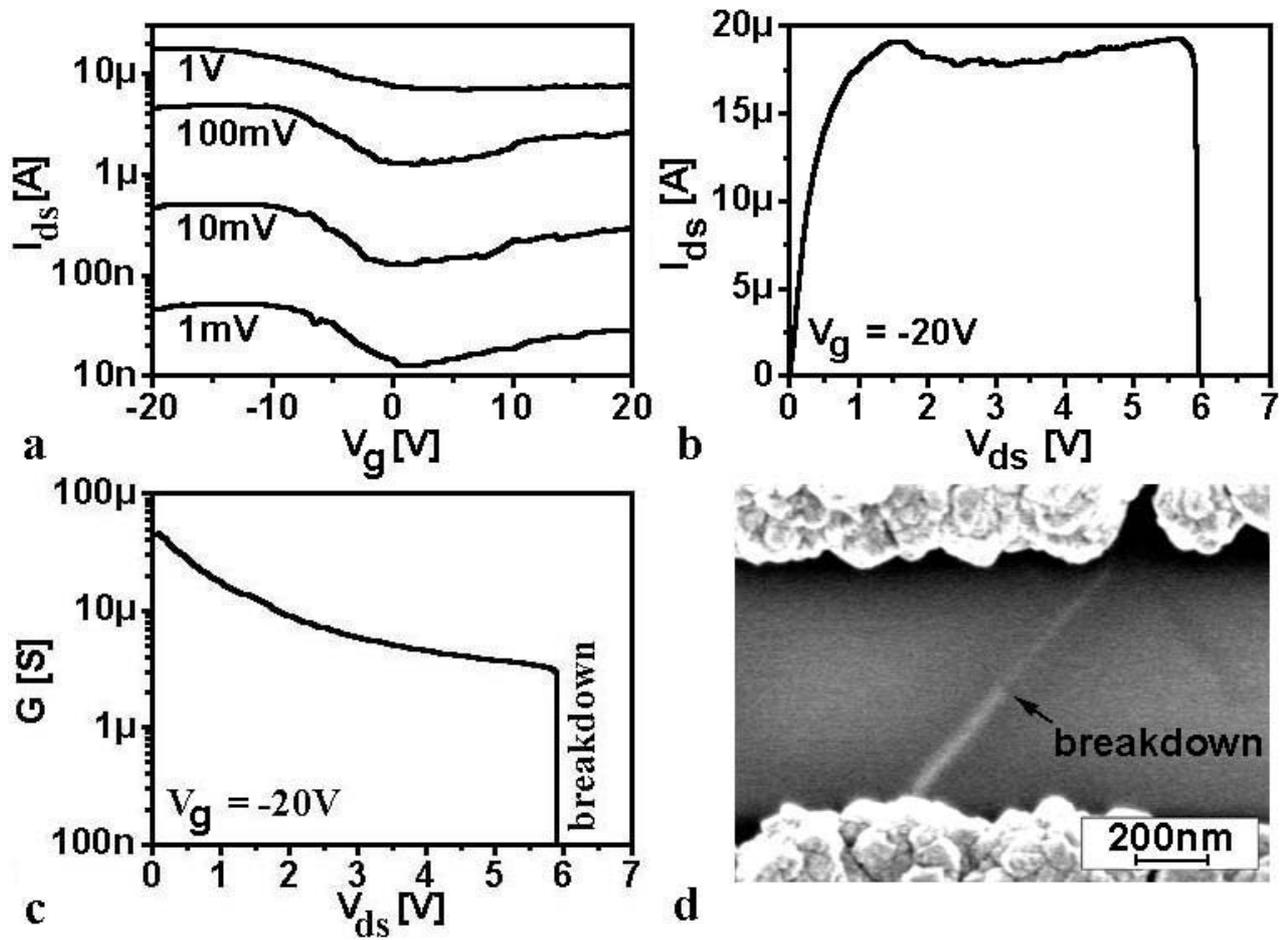

Figure 4

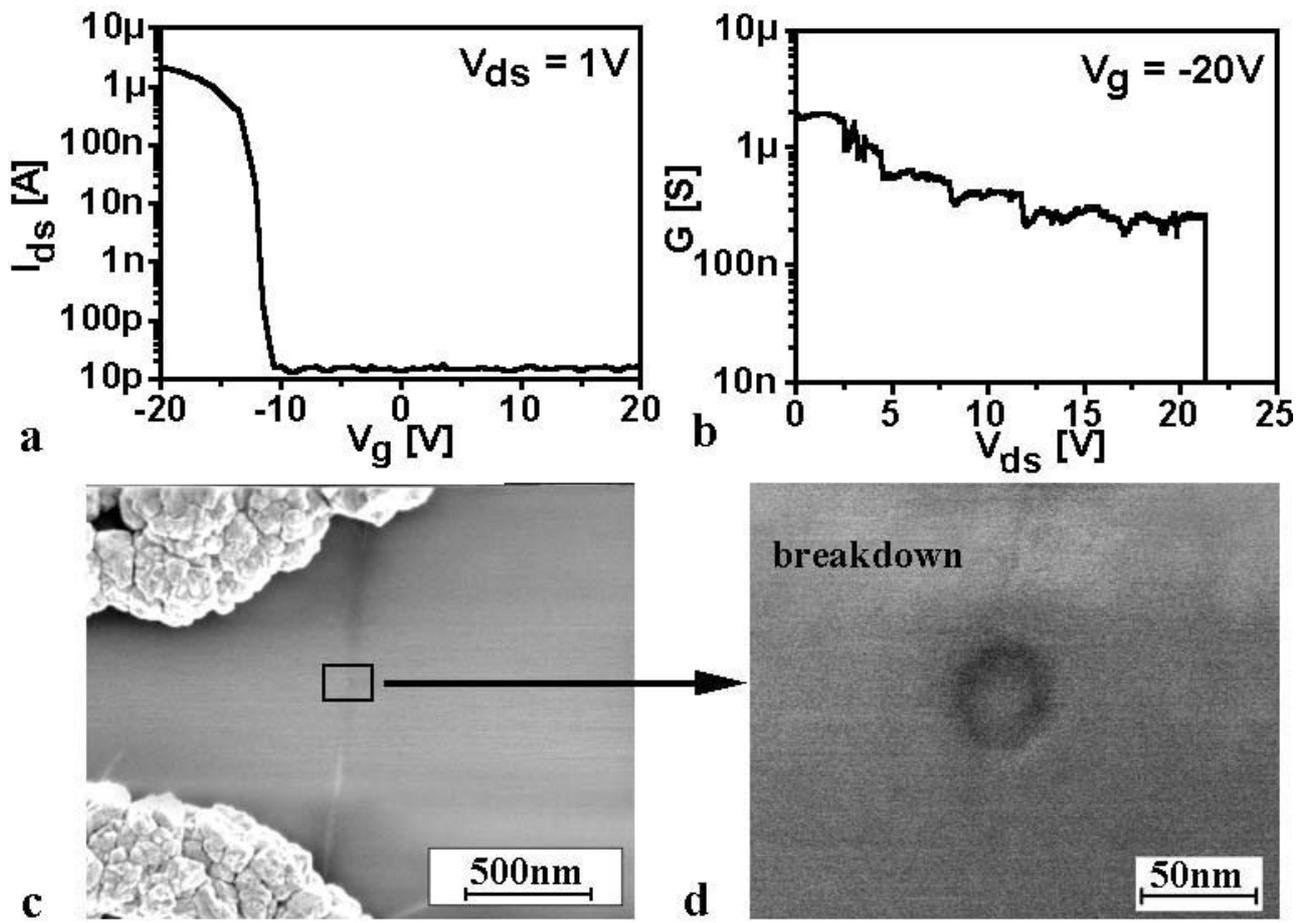

Figure 5

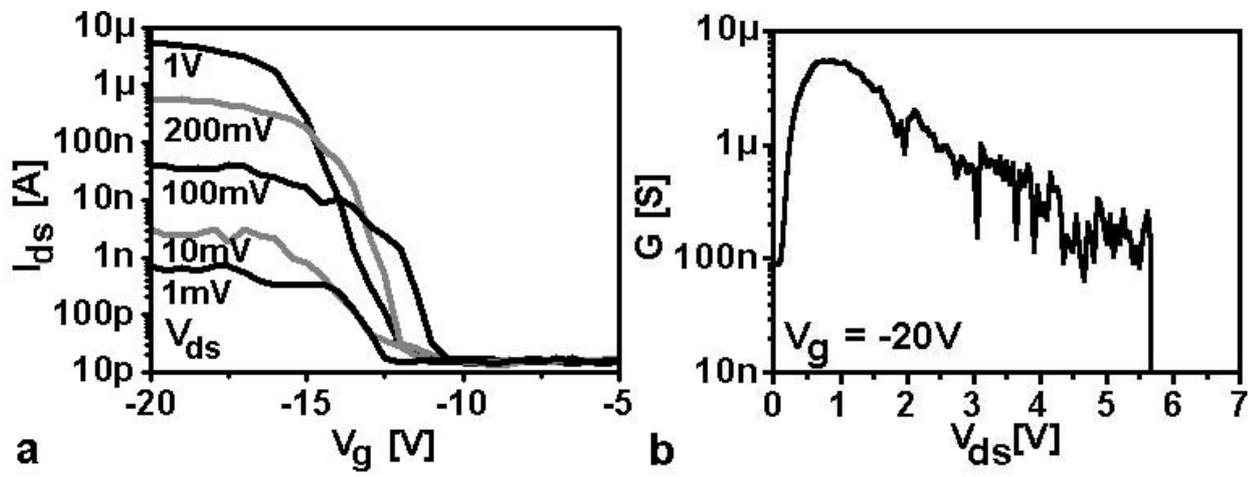

Figure 6

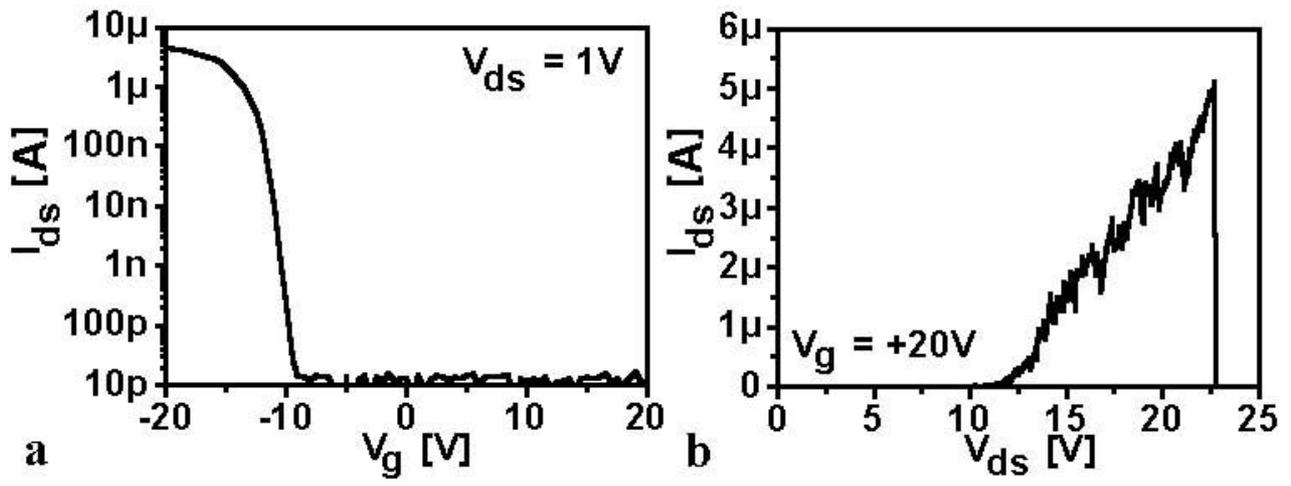

Figure 7

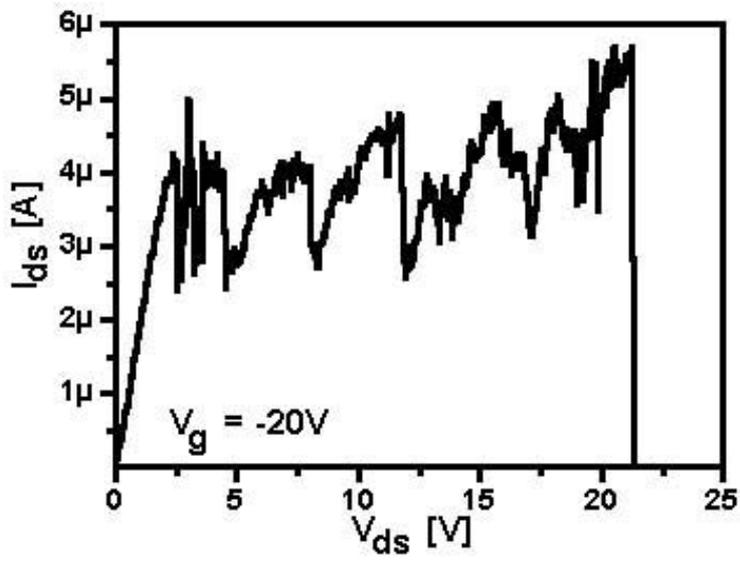

Figure 8

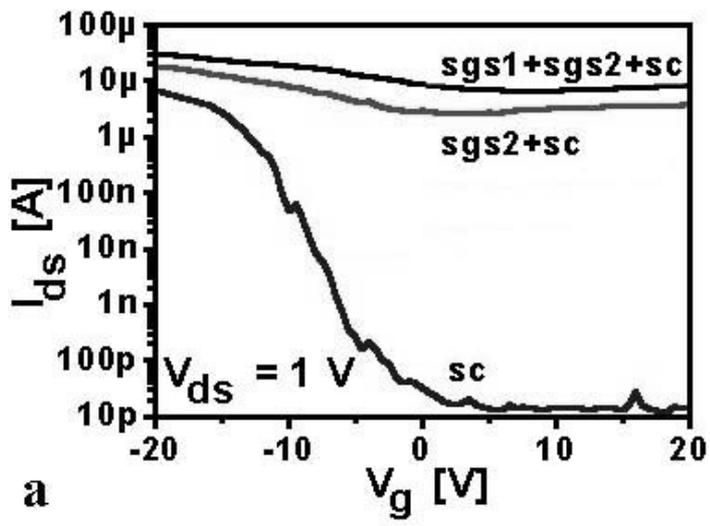

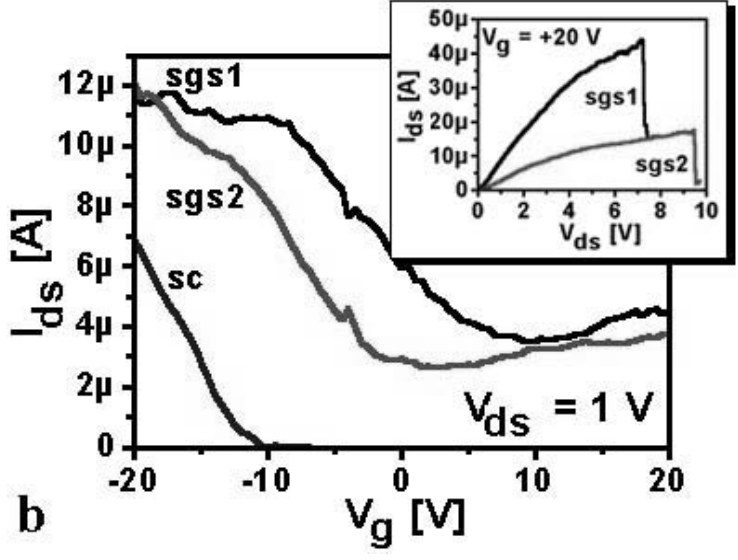

Figure 9